\documentclass[prl,twocolumn,unsortedaddress,superscriptaddress]{revtex4}
\usepackage{amsmath}  
\usepackage{amsbsy,amssymb,amsfonts}
\usepackage{graphicx}
\usepackage{color}
\usepackage{epsf}
\usepackage[utf8]{inputenc}
\usepackage{hyperref}
\usepackage{multirow}
\def\cm{$\rm cm^{-1}$}

\def\bravert{\egroup\,\vrule\,\bgroup}

{\catcode`\|=\active
  \gdef\Twoint#1{\left(\mathcode`\|"8000\let|\bravert {#1}\right)}}
{\catcode`\|=\active
  \gdef\Braket#1{\left<\mathcode`\|"8000\let|\bravert {#1}\right>}}

\newcommand{\beq}{\begin{equation}}
\newcommand{\eeq}{\end{equation}}
\newcommand{\beqa}{\begin{eqnarray}}
\newcommand{\eeqa}{\end{eqnarray}}
\newcommand{\bea}{\begin{array}}
\newcommand{\eea}{\end{array}}

\newcommand{\bef}{\begin{figure}}
\newcommand{\ef}{\end{figure}}
\newcommand{\bc}{\begin{center}}
\newcommand{\ec}{\end{center}}
\newcommand{\bt}{\begin{table}}
\newcommand{\et}{\end{table}}
\newcommand{\btb}{\begin{tabular}}
\newcommand{\etb}{\end{tabular}}

\def\molcas{$\cal M\kern-0.10em O\kern-0.15em L\kern-0.00em 
             C\kern-0.10em A\kern-0.05em S$}
\begin{document}

\vspace{2cm}
\title {{Electron Electric Dipole Moment and Hyperfine Interaction Constants for ThO}}

\vspace*{2cm}

\author{Timo Fleig}
\affiliation{Laboratoire de Chimie et Physique Quantiques,
             IRSAMC, Universit{\'e} Paul Sabatier Toulouse III,
             118 Route de Narbonne, 
             F-31062 Toulouse, France}

\author{Malaya K. Nayak}
\affiliation{Bhabha Atomic Research Centre,
             Trombay, Mumbai - 400085,
             India}

\date{\today}
\vspace*{1cm}
\begin{abstract}
A recently implemented relativistic four-component configuration interaction approach to study
${\cal{P}}$- and ${\cal{T}}$-odd interaction constants in atoms and molecules is employed
to determine the electron electric dipole moment effective electric field in the $\Omega=1$
first excited state of the ThO molecule. We obtain a value of 
$E_{\text{eff}} = 75.6 \left[\frac{\rm GV}{\rm cm}\right]$ with an estimated error bar of
$3$\% and $10$\% smaller than a previously reported result [arXiv:1308.0414 [physics.atom-ph]].
Using the same wavefunction model we obtain an excitation energy of $T_v^{\Omega=1} = 5329$
[\cm], in accord with the experimental value within $2$\%. In addition, we report the
implementation of the magnetic hyperfine interaction constant $A_{||}$ as an expectation
value, resulting in $A_{||} = -1335$ [MHz] for the $\Omega=1$ state in ThO. The smaller
effective electric field increases the previously measured upper bound to the electron
electric dipole moment interaction constant [arXiv:1310.7534v2 [physics.atom-ph]] and thus 
mildly mitigates constraints to possible extensions of the Standard Model of particle physics.
\end{abstract}

\maketitle
\clearpage

\section{Introduction}
\label{SEC:INTRO}

Polar diatomic molecules are promising complex systems 
\cite{Meyer_Bohn_Deskevich_2006,hudson_hinds_YbF2011} in search of the electric dipole moment (EDM) of 
the electron. The measurement of a non-zero permanent molecular EDM and the establishment of its origin 
in fundamental charge (${\cal{C}}$) and spatial parity (${\cal{P}}$) violating interactions 
\cite{EDMsNP_PospelovRitz2005}, for example inducing an electron EDM, would be a signature of New 
Physics beyond the Standard Model (SM) of elementary particles \cite{LeptonicCPviolation_RevModPhys2012}. 

The thorium monoxide (ThO) molecule has been found to be one of the most interesting candidates in this 
quest \cite{demille_ThO_JPB2011,demille_ThO_PRA2011}, among other aspects due to its very large internal 
effective electric field $E_{\rm eff}$ on unpaired electrons \cite{Meyer_Bohn_PRA2008}. 
Recently, the ACME collaboration reported an order of magnitude smaller upper bound to the electron
EDM interaction constant, $|d_e| < 8.7 \times 10^{-29}\,e\,$cm, obtained from a spin-precession measurement
on a pulse of ThO molecules \cite{ACME_ThO_ArXiV2013}. This upper bound for $d_e$ is principally 
determined from Eq. (\ref{EQ:DE_MASTER}),
\begin{equation}
 d_e = \frac{\Delta E_t}{E_{\rm eff}}
 \label{EQ:DE_MASTER}
\end{equation}
where $\Delta E_t$ is an upper bound to a measured transition energy and $E_{\rm eff}$ is the internal 
electric field at the position of the electron, giving rise to a dipolar interaction with the electron's 
postulated EDM. 
$E_{\rm eff}$ cannot be measured experimentally but has to be determined from a theoretical 
electronic-structure calculation on the respective molecule in the respective quantum state. 
On the one hand, large $E_{\rm eff}$ is a selection criterion for systems with a large EDM interaction 
and therefore holding promise for the electron EDM to actually be found. Second, an accurate value of 
$E_{\rm eff}$ is required for reliably constraining the parameter ranges of New Physics models going 
beyond the SM \cite{MSSMReloaded_Ellis2008}.
Great care is taken in assessing and minimizing errors in the determination of $\Delta E_t$.
It is obvious that the same care should be taken in the theoretical assessment of the effective
electric field $E_{\rm eff}$.
 
Previous calculations of $E_{\rm eff}$ in the relevant ``science'' state $\Omega=1$, arising mainly from 
the configuration $7s^1 6d^1$ (Th$^{2+}$ O$^{2-}$), have been reported by Meyer et al. 
\cite{Meyer_Bohn_PRA2008} and Skripnikov et al. \cite{Skripnikov_ThO_ArXiV2013}. In the former 
\cite{Meyer_Bohn_PRA2008} $E_{\rm eff} = 104 \left[\frac{\rm GV}{\rm cm}\right]$ has been obtained 
based on a semi-empirical model calculation which in part employs non-relativistic approximations and
a very limited set of electronic configurations. The latter study \cite{Skripnikov_ThO_ArXiV2013} determines 
$E_{\rm eff} = 84 \left[\frac{\rm GV}{\rm cm}\right]$ by means of a two-component relativistic 
single-reference coupled cluster (CC) approach.

The work presented in this paper is aimed at an accurate determination of $E_{\rm eff}$ in the $\Omega=1$
excited state of the ThO molecule, along with a clarification of which physical aspects play the
decisive role in obtaining this quantity reliably. In addition, we investigate the excitation energy
of the $\Omega=1$ state and the parallel hyperfine coupling constant which is regarded as a measure
of the quality of molecular wavefunctions employed in the determination of relativistic EDM
enhancement factors.

\section{Theory}
\label{SEC:THEORY}
\subsection{Electron EDM Hamiltonian}

The potential energy due to the electron EDM interaction in the molecule is determined as an
expectation value \cite{commins_EDM_1999} over the one-body Hamiltonian $\hat{H}_{\rm edm}$ as 
follows
\begin{eqnarray}
 \left< \sum\limits_{j=1}^N\, \hat{H}_{\rm edm}(j) \right>_{\psi} 
 \nonumber
   &=& -d_e \left< \gamma^0 \sum\limits_{j=1}^N\, {\bf{\Sigma}}_j {\bf\cdot{E}}_j \right>_{\psi} \\
   &=& \frac{2\imath c d_e}{e\hbar}\, \left< \gamma^0 \gamma^5\, \sum\limits_{j=1}^N\, \vec{p}_j\,^2 \right>_{\psi},
 \label{EQ:HEDM:KINERG}
\end{eqnarray}
where $N$ is the number of electrons, $\gamma$ are the standard Dirac matrices, $d_e$ is the 
electron EDM interaction constant,
$\bf{\Sigma} = \left(
                     \begin{array}{cc}
                      \vec{\sigma} & {\bf{0}} \\
                      {\bf{0}} & \vec{\sigma}
                     \end{array}
               \right)$ with $\vec{\sigma}$ the vector of Pauli spin matrices, and ${\bf{E}}_j$ 
is the electric field at the position of an electron ($j$). 
The wavefunction $\psi$ is determined from relativistic 4-component Configuration Interaction 
(CI) theory \cite{knecht_luciparII} for the $\psi_{\Omega=1}$ first electronically excited state 
of the ThO molecule, using the all-electron Dirac-Coulomb Hamiltonian. 
Details on the implementation of Eq. (\ref{EQ:HEDM:KINERG}) can be found in reference 
\onlinecite{fleig_nayak_eEDM2013}. The optimized coefficients in the linear expansion of 
$\psi$ in the basis of Slater determinants over 4-component Dirac spinors contain the 
approximate effects of electron correlations among the electrons explicitly treated in the CI 
expansion.

\subsection{Magnetic Hyperfine Interaction Constant}

Since the magnetic vector potential $\vec{A}$ due to the magnetic moment $\vec{\mu}_K$ of a nucleus $K$
at the position $\vec{r}$ of an electron in an atom is \cite{Fermi_hyperfine1929}
\begin{equation}
 \vec{A} = \frac{\vec{\mu}_K \times \vec{r}}{r^3}
\end{equation}
we can derive the parallel magnetic hyperfine interaction constant $A_{||}$ as the $z$ projection
of the expectation value of the corresponding perturbative Hamiltonian in Dirac theory
\begin{equation}
 A_{||} = \frac{\mu_{Th}}{I \Omega}\, 
             \left< \sum\limits_{i=1}^n\, \left( \frac{\vec{\alpha_i} \times \vec{r}_i}{r_i^3}
                               \right)_z \right>_{\psi}
\label{EQ:APARALLEL}
\end{equation}
where $I$ is the nuclear spin quantum number, $\alpha_k$ is a Hamiltonian-form Dirac matrix for particle $k$,
and $n$ is the number of electrons. Again, we evaluate Eq. (\ref{EQ:APARALLEL}) over the CI
wavefunction for the state $\psi_{\Omega=1}$.

\section{Application to ThO}
\label{SEC:APPL}
\subsection{Technical Details}
\label{SUBSEC:TECHN}

\subsubsection{General Setup}
\label{SUBSUBSEC:GENERAL}

For the determination of the nuclear hyperfine coupling constant we use the thorium isotope
$\rm {^{229}{T}h}$ for which the nuclear magnetic moment has been determined to be 
$\mu = 0.45 \mu_N$ \cite{Kazakov_229ThO_ArXiV2012}. Its nuclear spin quantum number is
$I = 5/2$. In all calculations the speed of light was set to 137.0359998 a.u.

\subsubsection{Atomic basis sets}
\label{SUBSUBSEC:BASIS}

Fully uncontracted atomic Gaussian basis sets of double-$\zeta$, triple-$\zeta$ and 
quadruple-$\zeta$ quality were used for the description of electronic shells. For thorium we
used Dyall's basis sets \cite{5fbasis-dyall,5fbasis-dyall-ccorr} and for oxygen the Dunning 
cc-pVNZ-DK sets \cite{Dunning_jcp_1989} with $N \in \{2,3,4\}$, as well as the aug-cc-pVTZ-DK 
set \cite{Dunning_jcp_1989}.
For thorium all $5d,5f,7s,6d$ correlating exponents were added to the basic n-tuple-$\zeta$
sets, amounting to \{$26s,23p,17d,13f,1g$\} uncontracted functions in case of double-$\zeta$
(in the following abbreviated as vDZ),
\{$33s,29p,20d,14f,4g,1h$\} in the case of triple-$\zeta$ (vTZ) and \{$37s,34p,26d,17f,8g,4h,1i$\}
in the case of quadruple-$\zeta$ (vQZ), respectively. The latter set in addition contains all
$6s,6p$ correlating exponents.

\subsubsection{Molecular Wavefunctions}
\label{SUBSUBSEC:WAVEFUNC}

Molecular calculations were carried out with a modified local version of the \verb+Dirac11+
program package \cite{DIRAC11}. Optimized molecular spinors have been obtained using the
Dirac-Coulomb Hamiltonian and all-electron four-component Hartree-Fock calculations. The basic 
model used for these open-shell calculations is based on an average-of-configuration Fock operator 
for two electrons in the Th ($7s,6d\delta$) Kramers pairs with all other ($88$) electrons
restricted to closed shells. This model, called (av.2in3), denoting an averaging with 2 electrons
in 3 Kramers pairs in the following, is appropriate for the region close to the equilibrium
bond distance of the molecule where the dominant configurations correspond to the system
Th$^{2+}$ O$^{2-}$ \cite{malmqvist_ThO_JCP2003}. The open-shell averaging ensures a balanced 
description of the low-lying electronic states of interest in this study. In a few models 
using the smallest basis set (double-$\zeta$) the $6d\pi\sigma$ and $7p$ shells of Th were 
included due to partial mixing with the Th $6d$ shell, defining (av.2in9). For the larger basis 
sets we have restricted the open-shell averaging to (av.2in3). 

\begin{figure}[h]
 \caption{\label{FIG:THO_GAS}
          Generalized Active Space models for ThO CI wavefunctions. The parameters $m, n, p$ 
          and $q$ are defined in the text and determine the occupation constraints of the 
          subspaces of Kramers-paired spinors. The molecular spinors are denoted according to 
          their principal atomic character. The space with $183-K$ virtual Kramers pairs (for 
          vTZ basis sets) is comprised by all canonical DCHF orbitals below an energy of $38$ 
          E$_H$.
         }


 \begin{center}
  \includegraphics[width=7.5cm,angle=0]{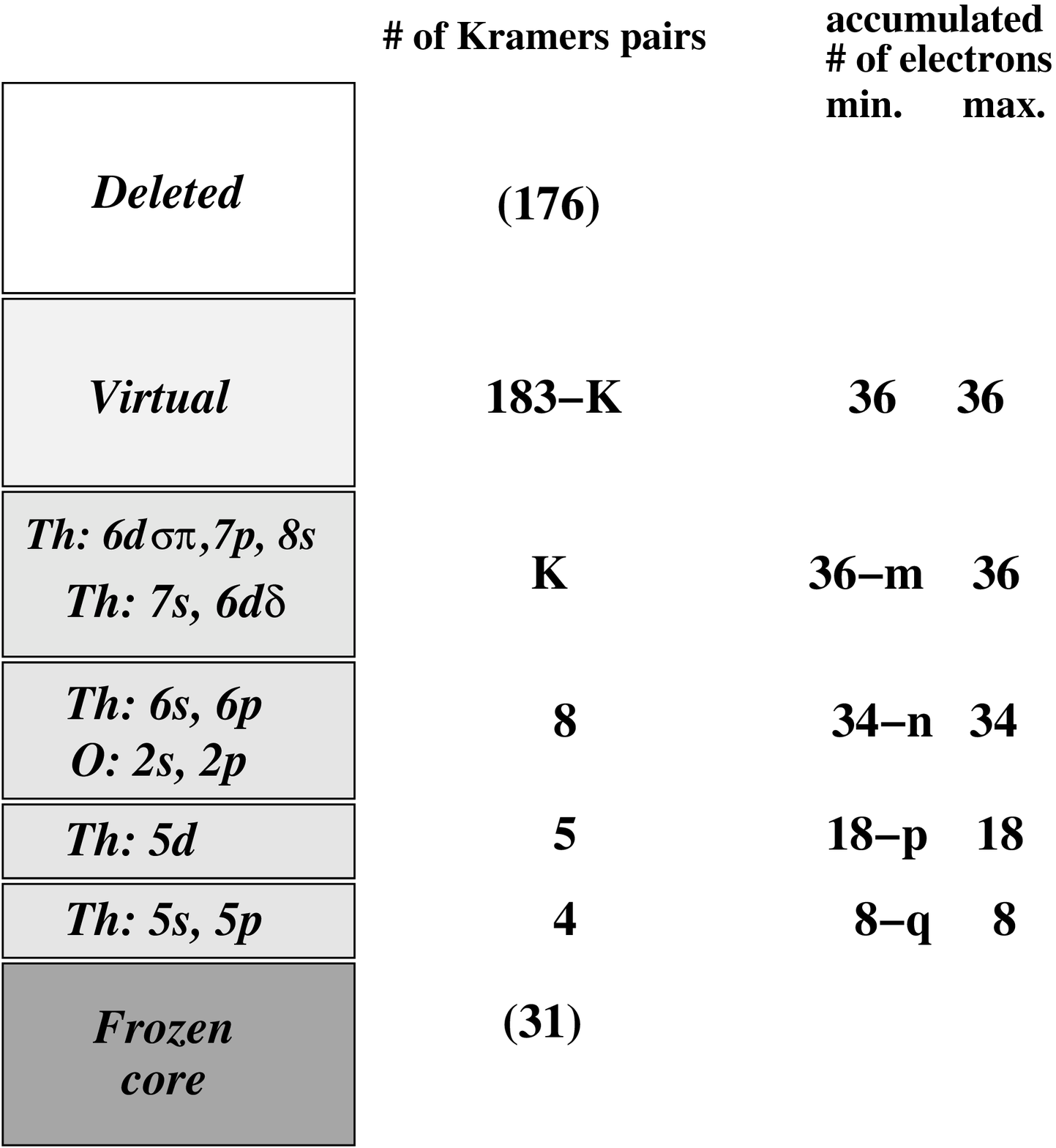}
 \end{center}
\end{figure}

We exploit a Generalized Active Space (GAS) concept for defining CI wavefunctions of varying
quality. Figure \ref{FIG:THO_GAS} shows the partitioning of the space of Kramers-paired
spinors into seven subspaces, five of which are active for excitations. Based on this


partitioning and four parameters ($m, n, p, q$) which define the accumulated occupation
constraints of the subspaces, we choose four different CI wavefunction models for our calculations:

\begin{center}
 \begin{tabular}{l|c}
  parameter values       & correlation model label     \\ \hline
  $m=2, n=2, p=0, q=0$   & MR$_K$-CISD(18)     \\
  $m=3, n=2, p=0, q=0$   & MR$_K$-CISDT(18)    \\ 
  $m=2, n=2, p=2, q=0$   & MR$_K$-CISD(28)     \\
  $m=2, n=2, p=2, q=2$   & MR$_K$-CISD(36)
 \end{tabular}
\end{center}

The parameter $K$ (see Fig. \ref{FIG:THO_GAS}) has been introduced to define variable active valence 
spinor spaces. $K=3$ includes only the Th $(7s,6d\delta)$ spinors in the fourth active space and thus 
comprises a minimal model for a balanced description of the
ground $\Omega=0$ state and the excited $\Omega\in\{1,2,3\}$ states which derive from ${^3\Delta}$
in the $\Lambda$-$S$ coupling picture. We have furthermore used $K=5$ which adds two $\pi$-type
spinors to the fourth space, $K=7$ adding another two $\pi$-type spinors, and finally $K=9$ and $K=10$
which adds energetically low-lying $\sigma$-type spinors to this space.

The different wavefunction models are in addition defined by the number of correlated electrons in
total (in parenthesis) and the included excitation ranks, where ``SDT'' stands for Single, Double,
and Triple excitations, as an example. The value of a parameter, e.g. $p=2$, denotes the maximum
hole rank of the respective active space. In that particular example all Slater determinants with
zero up to two holes in the Th $(5d)$ space would be included in the wavefunction expansion.

\subsection{Electronic-structure results}
\label{SUBSEC:SPECTR}

We first establish molecular wavefunctions which accurately describe the excitation energy of 
the $\Omega = 1$ electronic state. Table \ref{TAB:THO:BASIS} displays vertical excitation 
energies as a function of basis set. The vTZ set leads to a large correction of $-15$\%, 
whereas the vQZ set only yields another $-4$\%, less than $200$ \cm\ on the absolute. We 
therefore continue our investigation with the set of vTZ quality.

The next criterion we consider is the electronic shells included in the explicit treatment of 
dynamic electron correlation. In Table \ref{TAB:THO:ELECTRONS} we compile results from only 
$2$ correlated electrons (Th ($7s,6d$) shells) up to $38$ correlated electrons (Th 
($5s,5p,5d,6s,6p,6d,7s$), O ($2s,2p$) shells). Whereas correlations among the valence electrons 
of both atoms are seen to be important, core-valence and core-core correlations change the 
excitation energy by only $-3$\%, on the order of $-100$ \cm. 

As a third criterion we take the size and structure of the active spinor space into account. 
$X$ giving the number of Kramers pairs in the active space, the excitation energies for four 
different models are compared in Table \ref{TAB:THO:ACTIVE}. We observe that increasing the 
active space leads to non-negligible corrections. In particular the last step, from $X=7$ to 
$X=10$, where $\sigma$-type spinors are added to the active space, proves to be important. It 
is only at this level that the vertical excitation energy becomes satisfactorily accurate as
compared to the experimental value of $T_e = 5317$ \cm. Theoretical studies have shown 
\cite{malmqvist_ThO_JCP2003} that the difference between the equilibrium bond lengths in the 
$\Omega=0$ and the here considered $\Omega=1$ excited state amounts to only $0.03$ a.u. From 
this result we infer a small non-parallelity correction on the order of $-100$ \cm\ for our 
vertical excitation energy determined at the experimental minimum of the ground state 
potential-energy curve. 


\subsection{Electron EDM and Hyperfine Interaction Constants}
\label{SUBSEC:EDMCON}

Having studied the quality of the wavefunction in describing the excited state of relevance to 
the electron EDM measurement, we turn our attention to the determination of the effective 
electric field and the hyperfine interaction constants in this state.

The results in Table \ref{TAB:THO:BASIS} show that $E_{\text{eff}}$ is virtually insensitive

\begin{table}[h]

\caption{\label{TAB:THO:BASIS}
         Vertical excitation energy, effective electric field, and hyperfine constant
         at an internuclear distance of $R = 3.477$ a$_0$ for 
         $\Omega = 1$ using basis sets with increasing cardinal number and the wavefunction 
         model MR$_3$-CISD($18$)
        }

\begin{center}
\begin{tabular}{l|cccc}
 Basis set/CI Model     & $T_v$ [\cm] & $E_{\text{eff}} \left[\frac{\rm GV}{\rm cm}\right]$ & $A_{||}$ [MHz] \\ \hline
 vDZ/MR$_3$-CISD($18$)  &  $4535$     &  $80.8$  &  $-1283$       \\
 vTZ/MR$_3$-CISD($18$)  &  $3832$     &  $81.0$  &  $-1292$       \\
 vQZ/MR$_3$-CISD($18$)  &  $3643$     &  $80.7$  &  $-1298$       \\ \hline
 Exp. ($T_e$)\footnote{Reference \cite{malmqvist_ThO_JCP2003}}  & $5317$  &   &  
\end{tabular}

\end{center}
\end{table}

to the size of employed atomic basis sets for ThO. The hyperfine interaction constant $A_{||}$,
changes by hardly more than $1$\% in magnitude when increasing the basis set cardinal number 
from $2$ to $4$. Since increasing the cardinal number improves
these standard basis sets predominantly in the outer core and valence atomic regions, we have
also tested the effect of adding steep functions to the thorium atomic core. It is observed that
$E_{\text{eff}}$ changes no more than by $0.03$\%, depending on the respective Gaussian exponent
of the added function. In the work of Skripnikov et al. \cite{Skripnikov_ThO_ArXiV2013} diffuse
functions are used in the atomic basis set for the oxygen atom, since the dominant contribution
to the electronic states in question arise from Th$^{2+}$ O$^{2-}$ configurations \cite{malmqvist_ThO_JCP2003}, 
and therefore such diffuse functions may affect the present results. Replacing the oxygen vTZ
basis set by the aug-cc-pVTZ-DK set leads to a change of $E_{\text{eff}}$ of less than $0.01$\%
and reduces $T_v$ for $\Omega=1$ by only $32$ \cm. We therefore conclude that the vTZ basis set
on the oxygen atom without additional diffuse functions yield sufficiently accurate results.

It has been argued that the EDM effective electric field is predominantly a core property 
\cite{steimle_WC2011} and thus could be sensitive to the correlated movement of the inner-shell electrons 
of the respective heavy atom. Table \ref{TAB:THO:ELECTRONS} suggests that these correlation contributions 

\begin{table}[h]

\caption{\label{TAB:THO:ELECTRONS}
         Vertical excitation energy, effective electric field, and hyperfine constant
         at an internuclear distance of $R = 3.477$ a$_0$ for 
         $\Omega = 1$ correlating only the atomic valence shells down to including core-valence 
         and core-core correlation and using the vTZ basis sets
        }

\begin{center}
\begin{tabular}{l|cccc}
 CI Model            & $T_v$ [\cm] & $E_{\text{eff}} \left[\frac{\rm GV}{\rm cm}\right]$ & $A_{||}$ [MHz] \\ \hline
 MR-CISD($2$)        & $5929$      & $68.5$            & $-1264$         \\
 MR$_3$-CISD($18$)   & $3832$      & $81.0$            & $-1292$         \\
 MR$_3$-CISD($28$)   & $3752$      & $80.0$            & $-1297$         \\
 MR$_3$-CISD($36$)\footnote{Due to extreme computational demand the virtual cutoff is 5 a.u. here.}   
                     & $3742$      & $80.8$            & $-1287$         \\ \hline
 Exp. ($T_e$)\footnote{Reference \cite{malmqvist_ThO_JCP2003}}  & $5317$  &   &  
\end{tabular}

\end{center}
\end{table}

are negligible, at least for the ThO molecule. Whereas the inclusion of only the two valence electrons
is insufficient for any of the properties discussed here, $E_{\text{eff}}$ and the hyperfine coupling
constant prove to be sufficiently converged already at the valence level of $18$ correlated electrons.
As we have previously discussed for the HfF$^+$ molecular ion \cite{fleig_nayak_eEDM2013} this can be
rationalized as follows: 

In orbital perturbation theory, the first-order corrected expression for a given spinor $\varphi_i$
is
\begin{equation}
 \varphi_i \approx \varphi_i^{(0)} + \sum\limits_{k(\neq i)}\, 
           \frac{\left< \varphi_k^{(0)} | \hat{\cal{V}} | \varphi_i^{(0)} \right>}{\varepsilon_i^{(0)}
                                          - \varepsilon_k^{(0)}}\, \varphi_k^{(0)}
 \label{EQ:ORBPERT}
\end{equation}
where $\varphi_m^{(0)}$ is the $m$th unperturbed spinor, $\varepsilon_m^{(0)}$ the corresponding spinor
energy and $\hat{\cal{V}}$ is the electron correlation fluctuation potential. In the present case we
are interested in the change of the valence spinor $\varphi_{7s}$ which is the dominant contributor to
the electron EDM expectation value, Eq. (\ref{EQ:HEDM:KINERG}). If a spinor $\varphi_k^{(0)}$ is a(n)
(outer) core spinor, the energy difference in the denominator will become very large and thus the 
correlation contribution to the form of the valence spinor will be very small. 
As an example, we consider $\varphi_k^{(0)} = \varphi_{1s}$ and $\varphi_i^{(0)} = \varphi_{7s}$.
The fluctuation potential matrix element $\left< \varphi_k^{(0)} | \hat{\cal{V}} | \varphi_i^{(0)} \right>$
is on the order of $E_H$, whereas the energy denominator becomes $\varepsilon_i^{(0)} - \varepsilon_k^{(0)} 
\approx \left(-0.05 + 4058.5\right) E_H = 4058.45 E_H$. The perturbation coefficient of the $\varphi_{1s}$ 
spinor to the $\varphi_{7s}$ spinor is therefore strongly suppressed. This analysis is clearly 
confirmed by our results in Table \ref{TAB:THO:ELECTRONS}. Both $E_{\text{eff}}$ and $A_{||}$ are largely
unaffected by including more than the $18$ valence electrons in the explicit treatment of electron
correlation. Even without having taken into account the innermost electronic shells of the thorium
atom in the correlation treatment, we can conclude that such correlations, due to the increasing magnitude 
of the energy denominator in Eq. (\ref{EQ:ORBPERT}), will lead to negligble contributions to the 
wavefunction of valence spinors.

We now turn our attention to wavefunction models built from active spinor spaces of variable size,
the results for which are given in Table \ref{TAB:THO:ACTIVE}. The minimal active space (MR$_3$-CISD)

\begin{table}[h]

\caption{\label{TAB:THO:ACTIVE}
         Vertical excitation energy, effective electric field, and hyperfine constant
         at an internuclear distance of $R = 3.477$ a$_0$ for 
         $\Omega = 1$ using the vTZ basis set and varying active spinor spaces
        }

\begin{center}
\begin{tabular}{r|cccc}
 CI Model            & $T_v$ [\cm] & $E_{\text{eff}} \left[\frac{\rm GV}{\rm cm}\right]$ & $A_{||}$ [MHz] \\ \hline
      MR$_3$-CISD($18$)    & $3832$      & $81.0$            & $-1292$    \\
      MR$_5$-CISD($18$)    & $4054$      & $79.7$            & $-1291$    \\
      MR$_7$-CISD($18$)    & $4321$      & $80.1$            & $-1318$    \\
      MR$_{10}$-CISD($18$) & $5329$      & $75.6$            & $-1335$    \\ \hline
 Exp. ($T_e$)\footnote{Reference \cite{malmqvist_ThO_JCP2003}}  & $5317$  &   &  
\end{tabular}

\end{center}
\end{table}


yields a value of $E_{\text{eff}} = 81.0 \left[\frac{\rm GV}{\rm cm}\right]$ which is quite close to the
most elaborate result of Skripnikov et al. \cite{Skripnikov_ThO_ArXiV2013} which is 
$84.0 \left[\frac{\rm GV}{\rm cm}\right]$. As expected, the augmentation of the active spinor space
with spinors of different symmetry representation than $\varphi_{7s}$ does not significantly affect
$E_{\text{eff}}$ for $\Omega = 1$. However, upon including the $\sigma$-type spinors in the active
space, yielding the model MR$_{10}$-CISD($18$), we observe a strong drop of the effective electric
field to a value of $75.6 \left[\frac{\rm GV}{\rm cm}\right]$. This reduction is accompanied by a 
striking improvement of the vertical excitation energy of the $\Omega = 1$ state to a value in
excellent agreement with the experimental result, even after applying the non-parallelity correction
discussed earlier. The hyperfine constant exhibits an increase of slightly more than $3$\% in
magnitude when increasing the size of the active spinor space from MR$_{3}$ to MR$_{10}$ which
is a significantly larger change than those observed for different basis sets and varying number
of correlated electrons.

In view of the significant discrepancy of the present MR$_{10}$-CISD($18$) result for $E_{\text{eff}}$ 
from the value reported by Skripnikov et al. it is instructive to discuss the different wavefunction 
models that have led to these results. The fact that an increase of the size of the active space has 
a noticeable effect on a property of the molecule, here $E_{\text{eff}}$ and $A_{||}$, points to the 
importance of a certain class of higher excitations in the 
molecular wavefunction expansion, here suggesting a multi-reference character of the $\Omega = 1$ 
excited state. In order to gain more insight we have carried out additional studies including excitation
ranks higher that Doubles into the virtual spinor space, see Table \ref{TAB:THO:EXCITATIONS}, which

\begin{table}[h]

\caption{\label{TAB:THO:EXCITATIONS}
         Vertical excitation energy, effective electric field, and hyperfine constant
         at an internuclear distance of $R = 3.477$ a$_0$ for 
         $\Omega = 1$ using the vDZ basis set and varying maximum excitation rank
        }

\begin{center}
\begin{tabular}{l|cccc}
 CI Model           & $T_v$ [\cm] & $E_{\text{eff}} \left[\frac{\rm GV}{\rm cm}\right]$ & $A_{||}$ [MHz] \\ \hline
 MR$_3$-CISD($18$)  & $4535$      & $80.8$            & $-1283$    \\
 MR$_9$-CISD($18$)  & $5703$      & $73.8$            & $-1321$    \\
 MR$_3$-CISDT($18$) & $5166$      & $74.4$            & $-1340$  \\ \hline
 Exp. ($T_e$)\footnote{Reference \cite{malmqvist_ThO_JCP2003}}  & $5317$  &   &  
\end{tabular}

\end{center}
\end{table}

was computationally feasible for the smallest employed basis sets, vDZ. Again, the inclusion of
higher excitations leads to the characteristic drop in $E_{\text{eff}}$, here by $\approx -6.5$
$\left[\frac{\rm GV}{\rm cm}\right]$ (MR$_3$-CISDT($18$) relative to MR$_3$-CISD($18$)). Interestingly,
nearly the same decrease is observed when configurations including three particles in the virtual
space, $v^3$, are excluded from the CI expansion, model MR$_9$-CISD($18$), and the higher excitations are
restricted to a subset involving additional $\sigma$-type spinors, i.e., configurations of the type
$\sigma^1 v^2$. We note that the $\Omega=1$ excitation energy behaves in very similar way, with
higher excitations involving the enlarged active spinor space playing the major role and increasing
the excitation energy markedly (the overshooting value for MR$_9$-CISD($18$) is due to basis set
incompleteness, as the results in Table \ref{TAB:THO:BASIS} confirm). Also for $A_{||}$ we observe
a notable dependence on higher excitations in the wavefunction, to a large degree covered by the
augmentation of the active spinor space. Again, the most elaborate result of Skripnikov et al.
($-1296$ [MHz] in our units) is very close to our result using the insufficient active spinor space
($-1292$ [MHz]).

Apparently, it is configurations involving excitations into the augmenting active space spinors,
combined with correlation excitations into the virtual space that are of major importance for obtaining an
accurate value for $E_{\text{eff}}$, and not the higher excitations such as $v^3$ provided by the
Coupled Cluster (CC) model of Skripnikov et al. The CC model of the latter authors, in turn, is a
standard single-reference CC expansion which is applied to the property of an electronically excited
state which clearly exhibits a significant multi-reference character, as our results demonstrate.
In addition, Skripnikov et al. use ground-state ${^1\Sigma_0}$ spinors and the same state as the
Fermi vacuum for the CC expansion which biases the wavefunction towards this ground state. This is
confirmed by the too high excitation energy of the $\Omega=1$ state obtained by those authors 
($5741$ \cm) as compared to the experimental value of $5317$ \cm. In contrast to this result, our
best model vTZ/MR$_{10}$-CISD($18$) which is in addition based on configuration-averaged spinors 
yields an excitation energy in much better agreement with the
experimental value, and, furthermore, shows that the adequate description of all relevant physical 
effects results in a value of $E_{\text{eff}} = 75.6 \left[\frac{\rm GV}{\rm cm}\right]$, $10$\%
smaller than the earlier prediction of Skripnikov et al.

\section{Conclusion}
\label{SEC:SUMM}
By means of a careful study of effects due to basis sets, dynamic electron correlation, and
active spinor spaces within a rigorously relativistic all-electron four-component formalism,
we have determined the EDM effective electric field for the $\Omega=1$ first electronically
excited state at the experimental internuclear distance for the ThO molecular ground state.
We obtain a value of $E_{\text{eff}} = 75.6 \left[\frac{\rm GV}{\rm cm}\right]$ with an
estimated error bar of $3$\% and $10$\% smaller than the result previously reported by others
\cite{Skripnikov_ThO_ArXiV2013}. With the same wavefunction model we have obtained an
excitation energy for the $\Omega=1$ state of T$_v$ = $5329$ \cm, in excellent agreement with 
the experimental result which confirms the quality of the molecular wavefunctions employed
in the present study of ${\cal{P}}$ and ${\cal{T}}$ violating effects in ThO. The magnetic
hyperfine interaction constant is obtained to $A_{||} = -1335$ [MHz] using our most reliable 
wavefunction model.
Our detailed analysis shows that the ThO $\Omega=1$ state has a strong multi-reference nature
which must be adequately accounted for in electronic-structure theoretical treatments.

Our result for $E_{\text{eff}}$ has consequences for the recently determined upper bound to
the electron EDM interaction constant of $|d_e| < 8.7 \times 10^{-29}\,e\,$cm 
\cite{ACME_ThO_ArXiV2013}. This upper bound has been obtained based on 
$E_{\text{eff}} = 84 \left[\frac{\rm GV}{\rm cm}\right]$ \cite{Skripnikov_ThO_ArXiV2013}.
Due to Eq. (\ref{EQ:DE_MASTER}) a $10$\% smaller value for $E_{\text{eff}}$ requires a
corresponding adjustment of the upper bound for $d_e$ to a larger value. Consequently, the
electron EDM constraint to models extending the Standard Model of elementary particles is somewhat
attenuated.

In ongoing work we are studying potential energy curves for ThO along with dipole and transition 
dipole moments for relevant molecular electronic states. Furthermore, we are continuing with the
implementation of further operators of importance in the search for ${\cal{P}}$ and ${\cal{T}}$ 
violating effects in the universe, in particular the scalar-pseudoscalar ${\cal{P}}$,${\cal{T}}$-odd
electron-nucleon interaction.





\bibliographystyle{unsrt}
\newcommand{\Aa}[0]{Aa}








\end{document}